\documentclass[twocolumn,nofootinbib,prd,showpacs,aps,epsf,amsfonts,amssymb,amsbsy]{revtex4}
\usepackage{color}

\usepackage{graphics,bm}
\usepackage{epsfig}
\usepackage{graphicx}

\newcommand{\beq}{\begin{equation}}
\newcommand{\eeq}{\end{equation}}
\newcommand{\bqa}{\begin{eqnarray}}
\newcommand{\eqa}{\end{eqnarray}}

\def\square{\vcenter{\vbox{\hrule height.4pt
          \hbox{\vrule width.4pt height4pt
          \kern4pt\vrule width.3pt}\hrule height.4pt}}}

\begin{document}


\title{Two-color QCD in a strong magnetic field: The role of
the Polyakov loop}

\author{Arturo Amador}
\email{arturo.amador@ntnu.no}
\author{Jens O. Andersen}
\email{andersen@tf.phys.ntnu.no}
\affiliation{Department of Physics, 
Norwegian University of Science and Technology, H{\o}gskoleringen 5,
N-7491 Trondheim, Norway}

\date{\today}

\begin{abstract}
We study two-color QCD in a constant external magnetic 
backround at finite temperature using
the Polyakov-loop extended two-flavor two-color NJL model.
At $T=0$, the chiral condensate is calculated 
and it is found to increase  as a function of the
magnetic field $B$. In the chiral limit 
the deconfinement transition lies below
the chiral transition for nonzero magnetic fields $B$.
At the physical point, the two transitions seem to coincide
for field strengths up to $|qB|\approx 5M_{\pi}^2$, where
$M_{\pi}=140$ MeV,  
whereafter they split,
and the deconfinement transition takes place first.
The splitting between the
two increases as a function of $B$ in both the chiral limit and at the physical
point.
At the physical point, the transition temperature 
decreases slightly for magnetic fields up to $|qB|\approx 3 M_{\pi}^2$,
whereafter it increases monotonically.
In the chiral limit, this behavior is less pronounced.
This change of slope is absent in the NJL model where $T_c$ increases
for all values of $|qB|$.
In the range from zero magnetic field and $|qB|=20 M_{\pi}^2$, the
transition temperature for the chiral transition increases by
approximately 35 MeV, while the transition temperature for deconfinement
is essentially constant.
\end{abstract}

\maketitle

\section{Introduction}
The behavior of hadronic matter at finite temperature and density
in strong external 
magnetic fields has received a lot of attention for many years, 
see for example Ref.~\cite{book} for a very recent review.
The problem of strongly interacting matter in a strong magnetic background
arises in various contexts. For example, magnetars, which are a certain
type of neutron stars, have very strong magnetic fields of the order
of $10^{10}$ Tesla (T)~\cite{duncan}. Some of the properties of stars such as the
mass-radius relation are
determined by the equation of state. The determination of the bulk properties
of a Fermi gas in an external magnetic field is therefore important
for the understanding of these compact stellar objects.
Similarly, large magnetic fields, up
to the order of $eB\sim10^{14-16}$ T,
where $e$ is the electric charge of the pion,
are being generated in noncentral
heavy-collisions at the Relativistic Heavy-Ion Collider (RHIC) and
the Large-Hadron Collider (LHC)~\cite{mag1,mag2}. 
The presence of strong magnetic fields may be observed in these
experiments via the chiral magnetic effect.
This effect is basically the separation of charge
in a magnetic background due to the existence of topologically nontrivial
configurations in the deconfined phase of QCD~\cite{warringa}.
Finally, we mention that 
strong magnetic fields of the order $10^{14}-10^{19}$ T
may have been present in the early universe
during the strong and electroweak phase transitions~\cite{vaksa,olesen}.
The presence of a strong magnetic field 
at the electroweak phase transition 
may have implications for baryogenesis, i.~e. for the
generation of the baryon asymmetry
in the universe~\cite{laine,spanish}.  

Chiral symmetry of the QCD Lagrangian and the spontaneous breaking of this
symmetry in the QCD vacuum is an essential feature of the 
strong interactions.
At $T=0$, it is expected that a constant magnetic background
enhances chiral symmetry breaking if it is present already at $B=0$
or that it induces chiral symmetry breaking if the symmetry is intact
at $B=0$. This phenomenon is called magnetic catalysis
and has been discussed in 
Refs.~\cite{klevansky,klimenko,gusynin1,gusynin2,ebert22,shushp,werbos,janm}
in the context of the Nambu-Jona-Lasinio (NJL) model, chiral
perturbation theory, and QED (note however the recent 
paper~\cite{hidaka}
where the authors argue that effects from the neutral mesons might
show magnetic inhibition if $B$ is strong enough).
The basic mechanism is that neutral 
quark-antiquark pairs 
minimize their
energy by both aligning their magnetic moments along the direction of the
magnetic field~\cite{book}.
Magnetic catalysis was recently demonstrated 
on the lattice by Braguta et al~\cite{quench} in three-color quenched QCD
as well as by
Bali et al~\cite{budaleik,gunnar} 
in the context of three-color and 1+1+1
flavor QCD.
The results of~\cite{budaleik,gunnar}, which are 
for physical quark masses and extrapolated
to the continuum limit, are reproduced very well up to magnetic fields
of the order $eB\sim0.1$ (GeV)$^2$ in chiral perturbation 
theory~\cite{shushp,werbos} and up to  
$eB\sim0.25$ (GeV)$^2$ using the Polyakov-loop extended
NJL model (PNJL)~\cite{gatto1}.

Magnetic catalysis at $T=0$ gives rise to the expectation that
the critical temperature $T_c$ for the chiral transition is an increasing
function of the magnetic field $B$. Indeed, 
$\phi^4$-theory~\cite{duarte},
chiral perturbation 
theory~\cite{agassi,jensoa} (However, see also Ref.~\cite{fedo}), 
the NJL model~\cite{sid,pintotc},
the PNJL model~\cite{fuku,mizher,gatto1}, and the 
quark-meson (QM) model~\cite{rashid,skokov1,anders,pintotc} 
all predict this behavior (note however the 
recent paper where the authors use a $B$-dependent 
scale parameter for the Polyakov loop potential
to reproduce the lattice results~\cite{scoop}).
Furthermore, the PNJL model also predicts a modest split of approximately 2\%
between the chiral transition and the deconfinement transition, 
except for Ref.~\cite{gatto2}. In this case the split is of the order
10\% and is due to the effects of dimension 8 operators.
However, bag-model calculation~\cite{bag}, the
Polyakov-loop extended
QM calculation~\cite{mizher2}, and the large-$N_c$ calculation~\cite{largen} 
all predict a decreasing critical
temperature as a function of $B$. In Refs.~\cite{bag,mizher2}, 
it is probably
related to their treatment of vacuum fluctuations and related
renormalization issues.

Turning to lattice simulations, the picture
seems to be complicated as well. 
In Refs.~\cite{sanf,negro}, the lattice
simulations indicate that the chiral critical temperature is increasing
as a function of the magnetic field. In this case, the bare quark
masses used correspond to a pion mass in the range $M_{\pi}=200-480$ MeV, 
i. e. a very heavy pion. These results have been confirmed 
by Bali et al~\cite{budaleik,gunnar}.
However, for light quark masses that correspond to the physical
pion mass of $M_{\pi}=140$ MeV, their simulations
show a critical temperature which is a decreasing
function of the magnetic field $B$
The basic mechanism seems to be
that the magnetic catalysis at $T=0$ turns into inverse magnetic
catalysis~\cite{inverse1,inverse2} 
for temperatures around the critical temperature 
$T_c$~\cite{falk,gunnar2}.
The results suggest that the critical
temperature is a nontrivial function of the quark masses.

Two-color QCD is interesting for a number of reasons. 
The order parameter for the deconfinement transition depends on the
number of colors $N_c$. For $N_c=3$ it is known that the transition is
first order and for $N_c=2$ it is second order. Hence for $N_c=2$, one
expects universality and scaling close to the critical point.
For example, the critical exponents will be those of the 2-state Potts
model.
Moreover, in
contrast to three-color QCD, one can perform lattice simulations at
finite baryon chemical potential $\mu_B$. The reason is that due to the
special properties of the gauge group $SU(2)$, the infamous sign problem
is absent and thus importance sampling techniques can be used. 
Moreover, the physics at finite baryon chemical potential
is very different from its three-color counterpart:
Again due to the properties of the gauge group, two quarks can form
a color singlet and so diquarks are found in the spectrum
of the chirally broken phase.
The diquarks are bosons and finite baryon chemical potential
is then the physics of relativistic bosons and their condensation
at low temperature.
In the chiral limit, the Lagrangian of two-color two-flavor QCD has 
an $SU(4)$ symmetry. Since this group is isomorphic to $SO(6)$, chiral symmetry
breaking can be cast into the form $SO(6)\rightarrow SO(5)$.
The Goldstone modes are therefore contained in a single five-plet
with the usual three pions, a diquark and an antidiquark
as well.
Various aspects of the phase diagram of two-color QCD can be found
e. g. in Refs.~\cite{kond,kogut,kim,rotta,cea,simon,tilo,tomas1,jens,zhang,strod,kashiw,wiese}.

The problem of two-color QCD in a strong magnetic background was
first investigated on the lattice by Buidovidovich, Chernodub,
Luschevskaya, and Polikarpov~\cite{poli1,poli2} in the quenched 
approximation. 
Magnetic catalysis at $T=0$ has been verified 
and in in the chiral limit, the chiral condensate grows linearly 
for small values of $B$. This behavior is in qualitative agreement with
chiral perturbation theory.
Later, lattice simulations have been carried
out with dynamical fermions by Ilgenfritz, Kalinowski, 
M\"uller-Preussker, Petersson, and Schreiber
for $N_f=4$ with identical electric charges~\cite{peter}. 
We therefore make no comparision with the result presented here.
Their results seem to indicate that the condensate grows
linearly with $B$ in the chiral limit at $T=0$.
They also found that for all temperatures and 
fixed bare quark mass, the chiral condensate grows
with the magnetic field. This implies that the critical
temperature is an increasing function of the magnetic field.

In the present paper, we use the PNJL model to study two-color QCD
in a constant magnetic background $B$
at finite temperature and zero baryon chemical potential.
The article is organized as follows. In Sec.~II, we briefly discuss
the PNJL model in a magnetic field and the thermodynamic potential.
In Sec.~III, we present our numerical results and in Sec.~IV, we summarize
and conclude.

\section{PNJL model and thermodynamic potential }
In this section, we briefly discuss the two-flavor two-color PNJL model.
The Euclidean Lagrangian can be written as
\bqa
{\cal L}&=&{\cal L}_0+{\cal L}_1+{\cal L}_2\;,
\label{lag}
\eqa
where the various terms are
\bqa
{\cal L}_0&=&\bar{\psi}[i\gamma^{\mu}D_{\mu}-m_0]\psi\;,
\\ \nonumber
{\cal L}_1&=& 
G_1\left[(\bar{\psi}\psi)^2
+(\bar{\psi}i\gamma_5\psi)^2
+(\bar{\psi}{\boldsymbol \tau}\psi)^2
+(\bar{\psi}i\gamma_5{\boldsymbol \tau}\psi)^2
\right.\\ &&\left.
+|\overline{\psi^C}\sigma_2\tau_2\psi|^2
+|\overline{\psi^C}\gamma_5\sigma_2\tau_2\psi|^2
\right]\;,
\label{l1}
\\ \nonumber
{\cal L}_2&=& 
G_2\left[(\bar{\psi}\psi)^2
-(\bar{\psi}i\gamma_5\psi)^2
-(\bar{\psi}{\boldsymbol \tau}\psi)^2
+(\bar{\psi}i\gamma_5{\boldsymbol \tau}\psi)^2
\right.\\ &&\left.
-|\overline{\psi^C}\sigma_2\tau_2\psi|^2
+|\overline{\psi^C}\gamma_5\sigma_2\tau_2\psi|^2
\right]\;,
\label{l2}
\eqa
where the quark field $\psi$ is an isospin doublet
\bqa
\psi=
\left(\begin{array}{c}
u\\
d\\
\end{array}\right)\;.
\label{d0}
\eqa
The covariant derivative is 
$D_{\mu}=\partial_{\mu}-iqA_{\mu}-i\sigma_i A^{i}_{\mu}$, where
$A_{\mu}$ is the gauge field associated with $U(1)$ electromagnetism and 
$A^{i}_{\mu}$ is associated with $SU(2)$ color. 
The covariant derivative 
is diagonal in flavor space, $q={\rm diag}(2/3,-1/3)|e|$.
$\sigma_i$ ($i=1,2,3$) are the
Pauli matrices acting in color space, while 
$\tau_i$ are the
Pauli matrices acting in flavor space. 
$m_0$ is the mass matrix which is diagonal in flavor space
and contains the bare quark masses $m_u$ and $m_d$. In the following
we take $m_u=m_d$.
Moreover,
$\psi^C$ denotes the charge conjugate
of the Dirac spinor, $\psi^C=C\overline{\psi}^T$, where
$C=i\gamma^2\gamma^0$. 
$G_1$ and $G_2$ are coupling constants.
The interacting part ${\cal L}_1$ is invariant under global 
$U(4)=SU(4)\times U(1)_A$
transformations
while the 
${\cal L}_2$ is invariant under global $SU(4)$.
One sometimes writes $G_1=(1-\alpha)G$ and $G_2=\alpha G$ and so 
the parameter $\alpha$ determines the degree of 
$U(1)_A$ breaking. In the following we choose $\alpha={1\over2}$.

We next introduce the collective or auxilliary fields 
\bqa\nonumber
\sigma=-2G\bar{\psi}\psi\;,
\pi_i=-2G\bar{\psi}i\gamma_5\tau_i\psi\;,
\rho_i=-2G\bar{\psi}\tau_i\psi\;,
\label{coll2}
\\
\Delta=-2G\overline{\psi}i\gamma^5\tau_2\sigma_2\psi^C
\;,
\Delta_5=-2G
\overline{\psi}i\tau_2\sigma_2\psi^C\;,
\eqa
where $\sigma$, $\pi_i$, $\rho_i$, $\Delta$, and $\Delta_5$
have the 
quantum numbers of a scalar isoscalar, pseudoscalar isovector,
scalar isovector, scalar diquark, and psuedoscalar diquark, respectively.
The Lagrangian~(\ref{lag}) can then be written compactly as
\bqa\nonumber
{\cal L}
&=&
\bar{\psi}
\left[
i\gamma^{\mu}D_{\mu}-m_0
-\sigma-i\gamma^5\tau_i\pi_i
-\tau_i\rho_i
\right]\psi
\\ && \nonumber
+{1\over2}\left[\Delta^*\overline{\psi^C}i\gamma^5\tau_2\sigma_2\psi
+{\rm H.~c.}
+\Delta_5^*\overline{\psi^C}i\tau_2\sigma_2\psi+{\rm H.~c.}
\right]
\\ &&
-{1\over4G}\bigg[
\sigma^2+\pi^2_i
+\rho_i^2+|\Delta|^2+|\Delta_5|^2
\bigg]\;.
\label{coll}
\eqa
If we use the equation of motion for $\sigma$, $\pi_i$ 
$\tau_i$, $\Delta$, and $\Delta_5$
to eliminate
the auxilliary fields from the Lagrangian~(\ref{coll}), we obtain
the original Lagrangian~(\ref{lag}). 

In pure gauge theory, the Polyakov loop $\Phi$, which is the trace
of the Wilson line $l(x)=e^{i\int_0^{\beta}d\tau\,\sigma_iA_4^i(x,\tau)}$,
i.~e. 
$\Phi={1\over N_c}{\rm Tr}e^{i\int_0^{\beta}d\tau\,\sigma_iA_4^i(x,\tau)}$,
is an order parameter for deconfinement~\cite{svit}.
Under the center symmetry $Z_{N_c}$, it transforms as 
$\Phi\rightarrow e^{2\pi n/N_c}\Phi$, where $n=0,1,...N_c-1$.
For $N_c=2$, this is simply a change of sign and in two-color QCD
the Polyakov is purely real.
At low temperature, i. e. in the confined phase, we have $\Phi\approx0$
and in the deconfined phase, we have $\Phi\approx1$.
Note, however, that the Polyakov loop is only an approximate
order parameter in QCD with dynamical fermions.
In the PNJL model, a constant background temporal gauge field 
$A_4$ is introduced
via the covariant derivative in Eq.~(\ref{lag})~\cite{fukupol,megias}. 
In Polyakov gauge, the background field is diagonal in color space and,
$A_{4}^{i}=\theta\delta^{i,3}$, where $\theta$ is real. The Wilson line can then
be written as $l(x)={\rm diag}(e^{i\beta\theta},e^{-i\beta\theta})$
and the order parameter $\Phi$ 
reduces to
\bqa
\Phi&=&
\cos(\beta\theta)\;.
\eqa
In order to allow for a chiral condensate, 
we introduce a nonzero expectation value for the field
$\sigma$~\footnote{Since we consider the case of zero quark chemical
potential, the other collective fields have zero expectation value.} 
\bqa
\sigma&=&
-2G\langle\bar{\psi}\psi\rangle
+\tilde{\sigma}\;,
\eqa
where $\tilde{\sigma}$ is a quantum fluctuating fields
with vanishing expectation values.
To simplify the notation, we introduce the quantity $M$ which is defined by
\bqa
M&=&
m_0-2G\langle\bar{\psi}\psi\rangle\;. 
\label{mrho}
\eqa 
Note that the expectation value $M$ is assumed spacetime
independent in the remainder of this paper. 
Thus we ignore the possibility of inhomongeneous phases 
such as the Fulde-Ferrell-Larkin-Ovichinnikov phase as considered 
in~\cite{kim}.
Eq.~(\ref{coll}) is now
bilinear in the quark fields and we integrate them out exactly
by performing a Gaussian integral. This gives rise to an effective
action for the composite fields. In the mean-field approximation, we
neglect the fluctuations of the composite fields and 
the fermionic functional determinant reduces to 
\bqa\nonumber
\Omega_{\rm quark} &=&\frac{(M-m_{0})^{2}}{4 G}
-4N_c\int{d^3p\over(2\pi)^3}
\\ &&
\hspace{-1cm}\times
\bigg\{E_p+T\log\left[1+2\Phi e^{-\beta E_p}
\right.
\left.
+e^{-2\beta E_p}\right]\bigg\}\;.
\label{quarkc2} 
\eqa
where $E_p=\sqrt{p^2+M^2}$. 
Note that the integral involving $E_p$
is ultraviolet divergent and requires regularization. We will return to this
issue below. 

The interpretation of Eq.~(\ref{quarkc2}) is now as follows.
For $\Phi\approx0$, we have confinement and thus a
thermal part proportional to $T\log[1+e^{-2\beta E_p}]$, which
corresponds to an excitation of energy $2E_p$, i. e. a bound state.
Similarly, for $\Phi\approx1$, the thermal part is 
$T\log\left[1+2\Phi e^{-\beta E_p}+e^{-2\beta E_p}\right]=
2T\log[1+e^{-\beta E_p}]$ which is the thermal contribution from two degrees
of freedom each with energy $E_p$, i. e. 
the deconfined quark-antiquark pair.

The complete thermodynamic potential $\Omega$ is given by the sum
of the contributions from the quarks, $\Omega_{\rm quark}$ in 
Eq.~(\ref{quarkc2}) and a contribution from the gluons, $\Omega_{\rm gauge}$,
where~\cite{tomas1}
\bqa
\Omega_{\rm gauge}&=&
-bT\left[
24\Phi^2e^{-\beta a}+\log\left(1-\Phi^2\right)
\right]\;,
\label{gauge}
\eqa
where $a$ and $b$ are constants. This form is motivated by the 
lattice strong-coupling expansion~\cite{latfuk}.
In the pure gauge theory, we can find an explicit expression for
the value of the Polyakov loop, $|\Phi|=\sqrt{1-{1\over24}e^{\beta a}}$
and so $a=T_c\log24$. $\Phi$ goes to zero in a continuous manner
showing that the phase transition is second order in agreement
with universality arguments~\cite{svit}.

We next consider this system in a constant magnetic field $B$
along the $z$-axis. We do this by using the
covariant derivative 
$D_{\mu}=\partial_{\mu}-iqa_{\mu}-i\sigma_iA_{\mu}^i$,
where $a_{\mu}=\delta_{\mu,2}x_1B$ and 
$A_{\mu}^i=\delta^{i,3}\delta_{\mu,4}\theta$.
Note that the $SU(4)$ symmetry of the Lagrangian is broken in an 
external magnetic field due to the 
different electric charges of the $u$ and $d$ quarks.
The remaining symmetry is a $U(1)_A$ symmetry which corresponds to a
rotation of the $u$ and $d$ quarks by opposite angles~\cite{shushp}.
The chiral condensate breaks this Abelian symmetry 
and it gives rise to a single Goldstone boson, namely the neutral pion.

The energy eigenvalues of the Dirac equations 
are in this case given by 
\bqa
E_{m}&=&\sqrt{p_z^2+M^2+(2m+1-s)|q_fB|}\;,
\label{landauspec}
\eqa
where $M$ is the mass of the quark,
$s$ is the spin of the quark with electric charge $q_f$
and $m$ denotes the $m$th Landau level.
In Eq.~(\ref{quarkc2}), dispersion relation $E_p$ is now changed
to $E_m$ and the three-dimensional integral becomes a one-dimensional
integral and a sum of Landau levels $m$.
For a quark with charge $q_f$, we then make the replacements
\bqa
p^2&\rightarrow&p_z^2+(2m-1+s)|q_fB|\;,\\
\int{d^3p\over(2\pi)^3}&\rightarrow&
{|q_fB|\over2\pi}\sum_{m}\int{dp_z\over2\pi}\;,
\eqa
where the sum $m$ is over Landau levels and where the prefactor
${|q_fB|\over2\pi}$ takes into account the degeneracy of the Landau
levels.
The divergent term in 
Eq.~(\ref{quarkc2}) is denoted by $\Omega_{\rm quark}^{\rm div}$ and now
becomes
\bqa\nonumber
\Omega_{\rm quark}^{\rm div}
&=&-N_c\sum_{f,m,s}{|q_fB|\over\pi}\\ 
&&
\hspace{-1cm}
\times
\int{dp_z\over2\pi}\sqrt{p_z^2+M^2+(2m-1+s)|q_fB|}\;.
\label{defi}
\eqa
The integral over $p_z$ as well as the sum over Landau levels
$m$ in Eq.~(\ref{defi}) are divergent.
We will use zeta-function regularization and dimensional regularization
to regulate the divergences.
The integral is now generalized to $d=1-2\epsilon$ dimensions using
the formula
\bqa\nonumber
\int{d^d{p}\over(2\pi)^d}\sqrt{p^2+M_B^2}
&=&
-\left(
{e^{\gamma_E}\mu^2\over4\pi}\right)^{\epsilon}
{\Gamma(-{d+1\over2})\over(4\pi)^{d+1\over2}}
M_B^{d+1}
\;,
\\
\eqa
where $M_B^2=M^2+(2m+1-s)|q_fB|$ and
$\mu$ is the renormalization scale in the $\overline{\rm MS}$ 
renormalization scheme.
This yields
\bqa\nonumber
\Omega_{\rm quark}^{\rm div}
&=&-N_c\left({e^{\gamma_E}\mu^2\over4\pi}\right)^{\epsilon}
\Gamma(-1+\epsilon)\sum_{f,m,s}
{|q_fB|\over4\pi^2}M_B^{2-2\epsilon}\;,
\\&& 
\eqa
For each flavor $f$, the sum over $m$ and $s$ can be written as
\bqa\nonumber
\sum_{m,s}M_B^{2-2\epsilon}&=&
(2q_fB)^{1-2\epsilon}
\left[
\zeta(-1+\epsilon,x_f)-{1\over2}x_f^2\right]\;,
\\ &&
\eqa
where $\zeta(a,x)=\sum_{n=1}^{\infty}{1\over(a+n)^x}$ is the
Hurwitz zeta function and
$x_f={M^2\over2|q_fB|}$.
Expanding the Hurwitz zeta function in powers of
$\epsilon$, we obtain
\begin{widetext}
\bqa\nonumber
\Omega_{\rm quark}^{\rm div}
&=&
{N_c\over16\pi^2}
\sum_f
\left({\mu^2\over2|q_fB|}\right)^{\epsilon}
\left[
\left({2(q_fB)^2\over3}+M^4\right)\left({1\over\epsilon}+1\right)
-8(q_fB)^2\zeta^{(1,0)}(-1,x_f)-2|q_fB|M^2\log x_f
+{\cal O}(\epsilon)
\right]\;,
\\ &&
\label{divq}
\eqa
where $\zeta^{(1,0)}(-1,x_f)={d\over da}\zeta(a,x_f)|_{a=-1}$.
The first divergence, which is proportional to $(q_fB)^2$ can be removed
by wavefunction renormalization of the tree-level term ${1\over2}B^2$
in the free energy. This term is normally omitted since it is independent
of the other parameters of the theory.
The second divergent term, which is proportional to $M^4$
is the identical to the divergence that appears for $B=0$.
We can then add and subtract the term
\bqa
\int{d^dp\over(2\pi)^d}\sqrt{p^2+M^2}
&=&
\left({\mu\over M}\right)^{2\epsilon}{M^4\over2(4\pi)^2}
\left[{1\over\epsilon}+{3\over2}+{\cal O}(\epsilon)
\right]
\;,
\label{appear}
\eqa
to Eq.~(\ref{divq})
and take the limit $d\rightarrow3$ in the difference.
The divergence is now isolated in the integral on the left-hand-side of
Eq.~(\ref{appear}). We set $d=3$ here as well and 
and regulate it by imposing a sharp cutoff $\Lambda$ in the usual way.
Note that the UV cutoff $\Lambda$
is unrelated to the 
scale $\mu$ in dimensional regularization.
The quark thermodynamic potential then becomes
\bqa\nonumber
\Omega_{\rm quark}&=&{(M-m_0)^2\over4G}
-{N_c\over4\pi^2}
\left[\Lambda\sqrt{\Lambda^2+M^2}(2\Lambda^2+M^2)
+M^4\log{\Lambda+\sqrt{\Lambda^2+M^2}\over M}
\right]
\\ && \nonumber
-{N_c\over16\pi^2}\sum_{f}
(q_fB)^2\left[
8\zeta^{\prime}(-1,x_f)-4(x_f^2-x_f)\log x_f+2x_f^2
\right]
\\ &&
-\sum_{f,m,s}{|q_fB|T\over2\pi}\int_0^{\infty}{dp_z\over2\pi}
\log\bigg[1+2\Phi e^{-\beta E_m}
+e^{-2\beta E_m}\bigg]\;.
\label{effpotr}
\eqa
The complete thermodynamic potential in a constant magnetic
background is the sum of Eqs.~(\ref{gauge}) and~(\ref{effpotr})
and denoted by $\Omega$.
The values of $M$ and the Polyakov loop
$\Phi$ are found by minimizing $\Omega$
with respect to $M$ and $\Phi$, i. e. by solving the
gap equations
\bqa
{\partial\Omega\over\partial M}=0\;,
\hspace{1cm}
{\partial\Omega\over\partial\Phi}=0\;.
\eqa
Using Eqs.~(\ref{effpotr}), we obtain
\bqa\nonumber
\nonumber
\\
{(M-m_0)\over2G}
-{N_c\over2\pi^2}
\left[
\Lambda\sqrt{\Lambda^2+M^2}
+{\Lambda^2(2\Lambda^2+M^2)\over\sqrt{\Lambda^2+M^2}}
+2M^2\log{\Lambda+\sqrt{\Lambda^2+M^2}\over M}
\right.&& \\\left.\nonumber
+{M^4\over\Lambda^2+M^2+\Lambda\sqrt{\Lambda^2+M^2}}
-M^3\right]
-{N_c\over16\pi^2}\sum_{f}M|q_fB|
\bigg[8\zeta^{\prime}(0,x_f)
-4(2x_f-1)\log x_f+
8x_f
\bigg]
\\ \nonumber
+{N_c\over2\pi^2}\sum_{f,m,s}
{|q_fB|}\int_0^{\infty}dp_z
{M\over E_m}{\Phi e^{-\beta E_m}+e^{-2\beta E_m}
\over1+2\Phi e^{-\beta E_m}+e^{-2\beta E_m}}
&=&0\;, \\ 
b\Phi\left[
{1\over1-\Phi^2}-24e^{-\beta a}\right]
-
\sum_{f,m,s}{|q_fB|\over2\pi}
\int{dp_z\over2\pi}{E^{-\beta E_m}\over1+ 2\Phi e^{-\beta E_m}+e^{-2\beta E_m}}
&=& 
0\;.
\label{poleq}
\eqa
\end{widetext}
We notice in particular that $\Phi=0$ is the only solution 
to Eq.~(\ref{poleq}) at $T=0$ and the PNJL model then reduces to 
the NJL model.

\section{Numerical results}
The PNJL model has five different parameters, namely 
$G$, $m_0$, and $\Lambda$ in $\Omega_{\rm quark}$ and $a$ and $b$ in 
$\Omega_{\rm gauge}$. At $T=0$, $\Omega_{\rm gauge}=0$ and so
the PNJL model reduces to the NJL model.
We can therefore determine the parameters $G$, $m_0$, and $\Lambda$ separately.
For $N_c=3$, one normally choses an ultraviolet cutoff $\Lambda$
and tunes the parameters $m_0$ and $G$ such that one reproduces the
pion mass $M_{\pi}$ and the pion decay constant $F_{\pi}$ in the vacuum.
For $N_c=2$, we have no experiments to guide us and several different
choices have been made~\cite{rotta,he,tomas1}. We follow Ref.~\cite{tomas1}
that uses $N_c$ scaling arguments.
Since the pion decay constant is proportional to $\sqrt{N_c}$
and the pion mass is proportional to $N_c$, we simply rescale the 
three-color values by $\sqrt{2\over3}$ and $2\over3$, respectively.
This scaling gives $f_{\pi}=75.4$ MeV and $m_{\pi}=93.3$ MeV.
Note that we in the following 
refer to the case where $m_{\pi}=93.3$ MeV as the physical point.
In order to facilitate the comparison with similar plots 
in the literature where $N_c=3$, $|qB|$ is given in units of 
$M_{\pi}^2=(140{\rm MeV})^2$ and the scaling of the thermodynamic potential
is done by dividing by $F_{\pi}^4$, where $F_{\pi}=93$ MeV, i. e. 
the $N_c=3$ values of the pion mass and the pion decay constant.
With an ultraviolet cutoff of $\Lambda=657$ MeV, this gives
$G=7.23$ (GeV)$^{-2}$ and $m_0=5.4$ MeV at the physical point
and 
$G=7.00$ (GeV)$^{-2}$ and $m_0=0$ MeV in the chiral limit.
The parameter $a$ in $\Omega_{\rm gauge}$ is related to the critical 
temperature for deconfinement transition
in pure-glue QCD and reads $a=T_c\log24$.
In the pure gauge theory, $T_c$ is independent of the number of
colors $N_c$~\cite{pure} to a first approximation, see however 
Ref.~\cite{lucini} for a recent study for $4\leq N_c\leq8$. We will 
therefore use the critical temperature for pure-glue
QCD from lattice calculations with $N_c=3$, 
$T_c=270$ MeV. This yields $a=858.1$ MeV.
The parameter $b$ can be tuned so that 
the chiral transition takes place at approximately the same 
temperature as the deconfinement transition. 
Note, however, that the Polyakov loop is strictly not an order parameter
in the presence of dynamical fermions. It is a crossover
and the transition region is defined as a band in which $\Phi$ varies.
We define the transition region to be the temperatures where
$0.4<\Phi<0.6$
and the width of the band in the $B$--$T$ plane
tells one how fast the crossover is.
Nevertheless, we define a
deconfinement temperature by the condition $\Phi={1\over2}$.
This gives a curve in the $B$--$T$ plane and acts a useful guide to the eye.
The requirement $\Phi={1\over2}$ yields
$b=(210.5)^3$ (MeV)$^3$.
We finally point out that instead of using the criteria for
the deconfinement and chiral transition mentioned above, it is common
to define $T_c$ by the inflection point of the appropriate order parameter
as a function of $T$. We have performed a few sample calculations
to compare the two criteria. For both transitions the difference between
them is less than one percent.

\begin{figure}[htb]
\begin{center}
\setlength{\unitlength}{1mm}
\includegraphics[width=8.0cm]{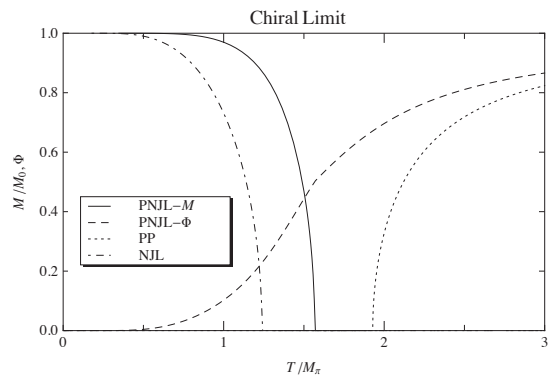}
\caption{Normalized constituent quark mass and Polyakov loop in the chiral limit
as a function of 
$T/M_{\pi}$, where $M_{\pi}=140$ MeV is the pion mass for $N_c=3$.
See main text for details.}
\label{tc0}
\end{center}
\end{figure}

In Fig.~\ref{tc0}, we show the 
normalized 
constituent quark mass $M/M_0$ (solid line), 
where $M_0$ is the quark mass at $T=0$,
and the 
Polyakov loop (dashed line)
in the chiral limit as a function of $T/M_{\pi}$, where
$M_{\pi}=140$ MeV is the pion mass in the vacuum
for $N_c=3$.
The quark mass vanishes at the temperature
at which $\Phi={1\over2}$. Thus for $|qB|=0$,
the two transitions take place at the same temperature as explained 
above~\footnote{In the chiral limit 
the normalized chiral condensate and the normalized
constituent quark mass are the same, cf. Eq.~(\ref{mrho}).}.
For comparison, we also plot the chiral condensate in the NJL model
(dotted line)
as well as the Polyakov loop in the pure-glue case (dash-dotted line), 
i.e. as derived
from the potential $\Omega_{\rm gauge}$ in Eq.~(\ref{gauge}).
The chiral transition in the chiral limit
is second order for $|qB|=0$ in the NJL as well as in the PNJL model.

\begin{figure}[htb]
\begin{center}
\setlength{\unitlength}{1mm}
\includegraphics[width=9.0cm]{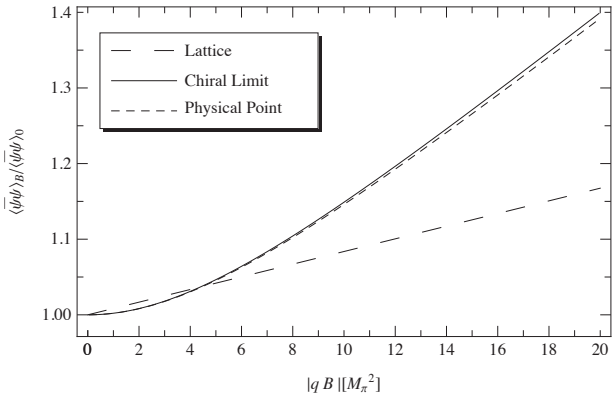}
\caption{The chiral condensate $\langle\bar{\psi}\psi\rangle_B$
normalized to the chiral condensate in the vacuum 
$\langle\bar{\psi}\psi\rangle_0$
as a function of magnetic field $|qB|$
in the chiral limit (solid line) and at the physical point (dashed line)
at zero temperature. Here $M_{\pi}=140$ MeV
is the pion mass for $N_c=3$.
Lattice results for pure-glue $SU(2)$ from Ref.~\cite{poli1}
is shown for comparison (long-dashed line).}
\label{mvsbch}
\end{center}
\end{figure}

In Fig.~\ref{mvsbch}, we show the chiral condensate 
$\langle\bar{\psi}\psi\rangle_B$
normalized to the chiral condensate in the vacuum 
$\langle\bar{\psi}\psi\rangle_0$
as a function
of $|qB|$ in the chiral limit and at the physical point
in the vacuum i.e. $T=\mu_B=0$ in the PNJL model. 
Note that here and in the following $|qB|$ is measured in units of
the pion mass for $N_c=3$, i.~e. $M_{\pi}=140$ MeV.
Since the effects
of the Polyakov vanish in the vacuum this is also the 
prediction of the NJL model. 
$\langle\bar{\psi}\psi\rangle_B$
is a monotonically
increasing function of $|qB|$ and the system exhibits magnetic catalysis.
In the NJL model and in the PNJL model at zero temperature, it is known
that the chiral condensate grows quadritically with the field for small
$|qB|$ in the chiral limit~\cite{klevansky,ebert22}.
This is in contrast to chiral perturbation theory where the 
dependence is linear. In Ref.~\cite{poli1} the authors investigate
the chiral condensate as a function of the magnetic field $B$
for $SU(2)$ gauge theory using lattice simulations in the quenched 
approximation. The authors found that the linear behavior found in chiral
perturbation theory can be described qualitatively 
by the function $\Sigma(B)=\Sigma_0\left(1+{{|qB|\over\Lambda_B^2}}\right)$,
where $\Sigma_0$ and $\Lambda_B$ are fitting parameters.
The parameters depend on the lattice parameters and we are using the
value $\Lambda_B=1.53$ GeV, which corresponds to their largest lattice
and their smallest lattice spacing $a$.
The result is shown as the long-dashed line in Fig.~\ref{mvsbch}
and is seen to agree reasonably well for magnetic field
up to $|qB|\approx 6M_{\pi}^2$

\begin{figure}[htb]
\begin{center}
\setlength{\unitlength}{1mm}
\includegraphics[width=9.0cm]{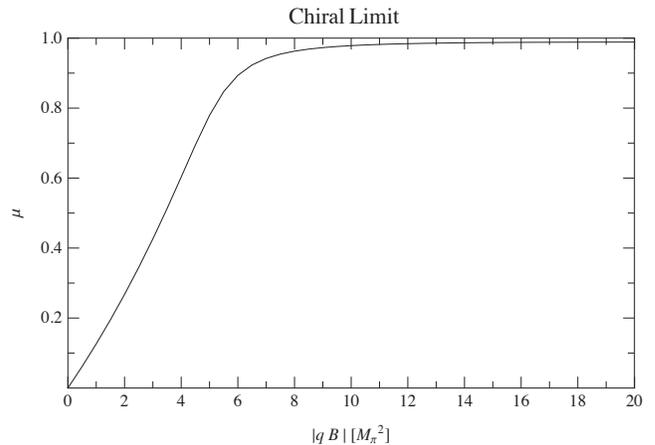}
\caption{The polarization $\mu$ as a function of $|qB|$
in the chiral limit and at zero temperature. Here
$M_{\pi}=140$ MeV is the pion mass for $N_c=3$.}
\label{pol}
\end{center}
\end{figure}

We next consider the magnetic moment for a fermion of flavor $f$. 
In terms of the spin operator 
$\Sigma^{\mu\nu}={1\over2i}
(\gamma^{\mu}\gamma^{\nu}-\gamma^{\nu}\gamma^{\mu})$, it is defined by
$\langle\bar{\psi}_f\Sigma^{\mu\nu}\psi_f\rangle$.
In the case of constant magnetic field in the $z$-direction, only
$\Sigma_f\equiv\Sigma^{12}$ is nonzero~\cite{frasca}.
Using the properties of the $\gamma$-matrices, it can be shown that only
the lowest Landau level (LLL) contributes to the expectation value
of $\Sigma_f$ and
reads
\bqa
\langle\bar{\psi}_f\Sigma_f\psi_f\rangle_B
&=&
\langle\bar{\psi}_f\psi_f\rangle^{\rm LLL}_B\;,
\eqa
where the superscript indicates that we include only the lowest Landau level.
We then define the polarization by
\bqa\nonumber
\mu_f&=&
{\langle\bar{\psi}_f
\Sigma_f\psi_f\rangle_B\over\langle\bar{\psi}_f\psi_f\rangle_B}
\\
&=&
1-{\langle\bar{\psi}_f\psi_f\rangle^{\rm HLL}_B
\over\langle\bar{\psi}_f\psi_f\rangle_B}\;,
\eqa
where the superscript $\rm HLL$ indicates that we have include only
the higher Landau levels.
In Fig.~\ref{pol}, we show the polarization $\mu={1\over2}(\mu_u+\mu_d)$
at $T=0$ and in the chiral
limit as a function of $|qB|$. As expected, the polarization saturates
for large magnetic fields to $\mu^{\infty}=1$.
In this limit, the fermions in the higher Landau levels effectively
become very heavy (cf. Eq.~(\ref{landauspec})), they  decouple
and the LLL dominates the physics. In this limit all the fermions are
in the LLL and their spin is pointing in the same direction.
The ratio of the mass of the fermions in the LLL and those
in the HLL is essentially given by the dimensionless ratio
$M^2/|qB|$. This ratio changes where the curve is steep and levels off
for large values of $|qB|$.

\begin{figure}[htb]
\begin{center}
\setlength{\unitlength}{1mm}
\includegraphics[width=8.0cm]{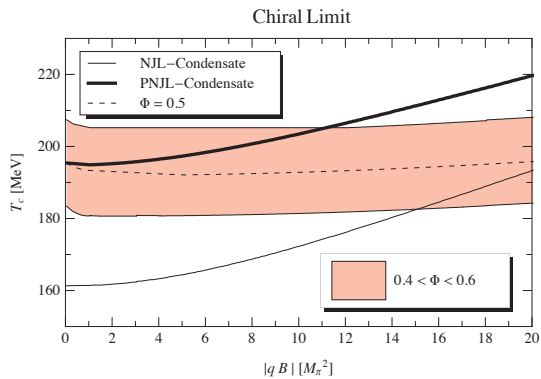}
\caption{Critical temperature for the chiral transition
(solid line) and critical temperature for the 
deconfinement transition (dashed line)
as functions of magnetic field $|qB|$ in the chiral
limit for the PNJL model. The band is defined by
$0.4<\Phi<0.6$.
We also show the critical temperature
for the chiral transition (thin line) in the NJL model. 
Here $M_{\pi}=140$ MeV is the pion mass for $N_c=3$.}
\label{tctrans}
\end{center}
\end{figure}

In Fig.~\ref{tctrans}, we show the critical temperature for the 
chiral transition (solid line) and the critical temperature
for the deconfinement transition (dashed line)
as functions of the magnetic field
in the chiral limit for the PNJL model. 
The band is defined by $0.4<\Phi<0.6$ and shows the transition region for
the deconfinement transition.
We also show the critical temperature in the NJL model for comparison.
The parameter $b$ 
in Eq.~(\ref{gauge}) has been tuned
such that the two transitions coincide for $B=0$.
We note that there is a splitting between the two transitions 
and that $T_c$ for the deconfinement is always lower than $T_c$
for the chiral transition and that the gap increases as a function of $B$.
Thus, there should be a phase 
in which matter is deconfined and chiral symmetry is broken.
The splitting was also observed in 
Ref.~\cite{gatto1,fuku,gatto2,mizher2,inverse2,calle},
where the authors coupled the Polyakov loop to linear sigma model with quarks
with $N_c=3$ colors.
We first note that
$T_c$ is decreasing ever so slightly from $|qB|=0$ to $|qB|\approx M_{\pi}^2$
and then increasing again. This is in contrast to the NJL model 
where $T_c$ is monotonically increasing as a function of $|qB|$.
We discuss this further below.

Moreover, while $T_c$ for the chiral transition increases by more than 20\%
from $B=0$ to $|qB|=20M_{\pi}^2$, $T_d$ for the deconfinement transition
is hardly affected. The width of the band is approximately 30 MeV. 
In contrast, the lattice simulations for $N_c=2$ reported in 
Ref.~\cite{peter} indicate that the
critical temperature for deconfinement coincide with
that of the chiral transition.
Finally, we note that the determination of $a$ and $b$ has changed
the critical temperature for the chiral transition dramatically.
The increase of $T_c$ at $B=0$ is approximately $35$ MeV and is fairly
constant up to $|qB|=20$ MeV. In order to compare the chiral transition
at finite magnetic field in the NJL and PNJL model, i. e. the effects
of the Polyakov loop, there might be other ways of determining 
$a$ and $b$. For example, one could force the deconfinement transition 
and the chiral transition in the PNJL model to take place at the 
same temperature for $B=0$ and force it to coincide with the
chiral transition in the NJL model as well.
This way of determining  the parameters in $\Omega_{\rm gauge}$
would not require the input from lattice simulations.

\begin{figure}[htb]
\begin{center}
\setlength{\unitlength}{1mm}
\includegraphics[width=9.0cm]{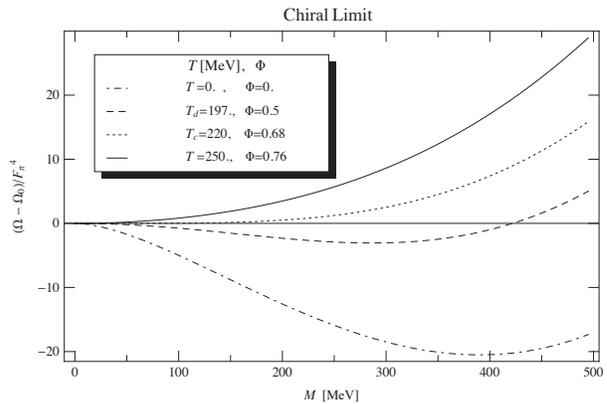}
\caption{Thermodynamic potential $\Omega-\Omega_0$ divided by $F_{\pi}^4$
as a function of $M$ in the chiral limit
for four different temperatures and $|qB|=20M^2_{\pi}$.
Here $M_{\pi}=140$ MeV and $F_{\pi}=93$ MeV are the values for the pion mass
and the pion decay constant for $N_c=3$.
See main text for details.}
\label{potvscondfixB}
\end{center}
\end{figure}

In Fig.~\ref{potvscondfixB}, we show the thermodynamic potential
$\Omega-\Omega_0$ divided by $F_{\pi}^4$ 
in the chiral limit for four different temperatures
and $|qB|=20M_{\pi}^2$.
For each temperature, we also give the value of the Polyakov loop.
The critical temperature for the chiral transition is $T_c=220$ MeV
and from the long-dashed line we see that transition is second order.
Since the value of the Polyakov loop for $T=220$ MeV is $\Phi=0.68$,
we conclude that the deconfinement transition has already taken place.

\begin{figure}[htb]
\begin{center}
\setlength{\unitlength}{1mm}
\includegraphics[width=8.0cm]{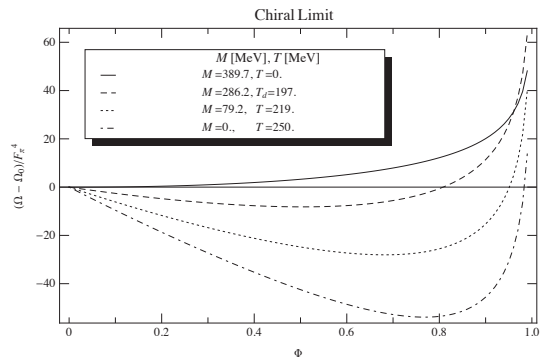}
\caption{Thermodynamic potential $\Omega-\Omega_0$ divided by $F_{\pi}^4$
as a function of $\Phi$ in the chiral limit 
for four different temperatures and $|qB|=20M^2_{\pi}$.
Here $M_{\pi}=140$ MeV and $F_{\pi}=93$ MeV are the values for the pion mass
and the pion decay constant for $N_c=3$.
See main text for details.}
\label{potvspl}
\end{center}
\end{figure}

In Fig.~\ref{potvspl}, we show the
thermodynamic potential $\Omega-\Omega_0$ divided by $F_{\pi}^4$
as a function of $\Phi$ in the chiral limit 
for four different temperatures and $|qB|=20M^2_{\pi}$. 
For each temperature, we also give the value of the chiral condensate $M$.
At $T=197$ MeV, we find that the minimum of the effective potential
occurs for
$\Phi={1\over2}$ which defines the critical temperature
for the deconfinement transition in the chiral limit.
At this temperature, $M=286.2$ MeV and so we are still in the 
chirally broken phase.
For $T=250$ MeV, the chiral condensate is vanishing and so 
we are in the chirally symmetric deconfined phase.
In the chiral limit, the chiral transition is always second order.

\begin{figure}[htb]
\begin{center}
\setlength{\unitlength}{1mm}
\includegraphics[width=9.0cm]{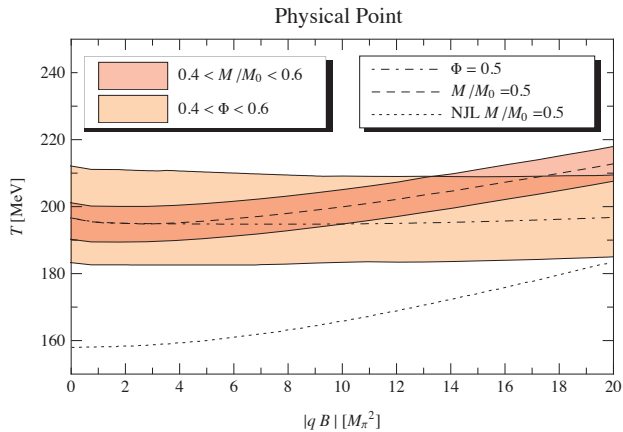}
\caption{Transition region at the physical point as a function of $|qB|$.
Here $M_{\pi}=140$ MeV is the value for the pion mass
for $N_c=3$.
For comparison, the
dashed line shows the pseudocritical temperature
in the NJL model.
See main text for details.}
\label{physpoint}
\end{center}
\end{figure}

\begin{widetext}

\begin{figure}[htb]
\setlength{\unitlength}{1mm}
\includegraphics[width=17.0cm]{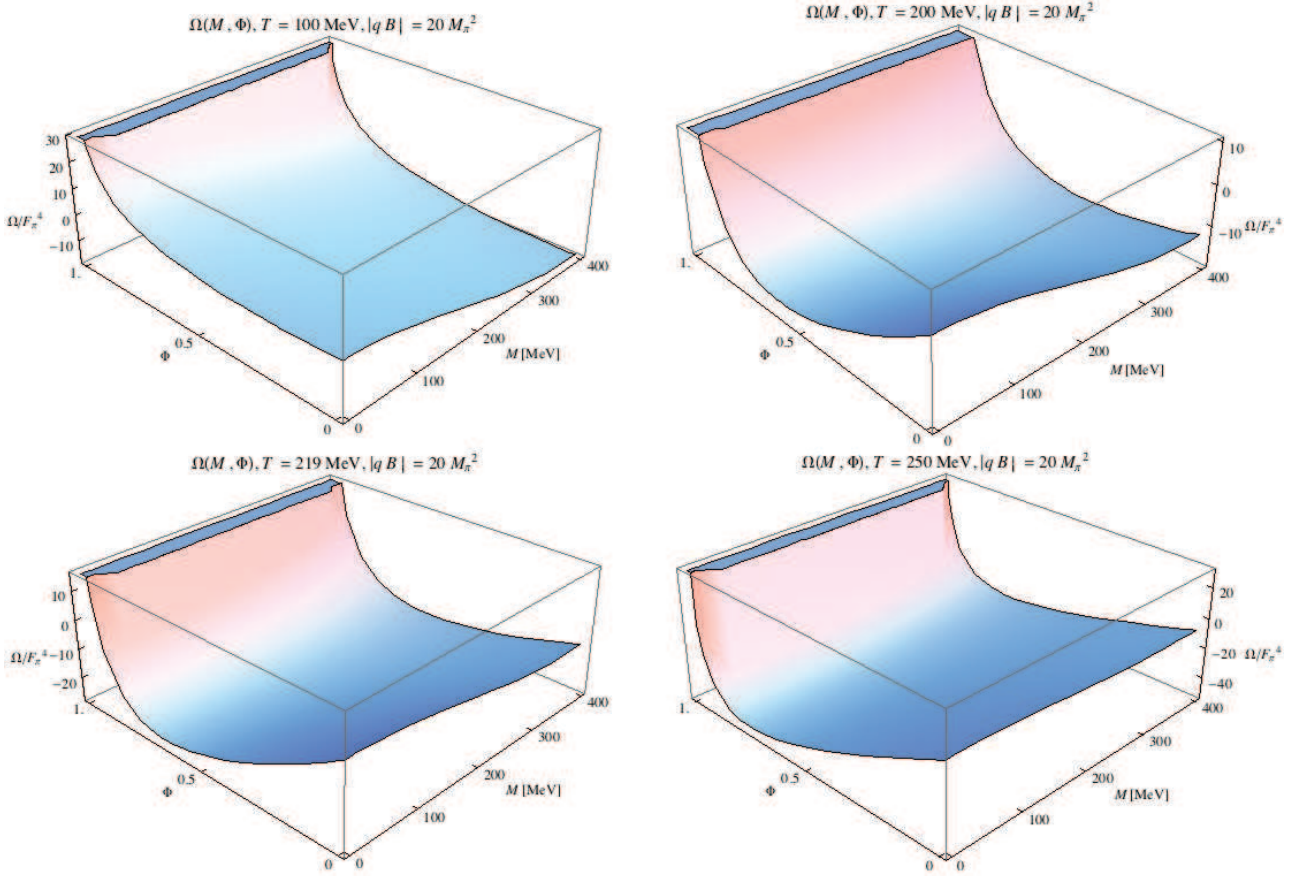}
\caption{The normalized thermodynamic potential 
$(\Omega-\Omega_0)/F_{\pi}^4$
in the chiral limit
as a function of $M$ and $\Phi$ for four different temperatures and
$|qB|=20 M_{\pi}$. From left to right $T=100$ MeV, $T=200$, Mev, 
$T=219$ MeV, and $T=250$ MeV. 
Here $M_{\pi}=140$ MeV and $F_{\pi}=93$ MeV are the values for the pion mass
and the pion decay constant for $N_c=3$.
See main text for details.}
\label{s1}
\end{figure}

\end{widetext}

At the physical point, the chiral transition is a crossover and there is
no well-defined critical temperature $T_c$ at which the chiral condensate
vanishes. However, one can define a pseudo-critical temperature by
the inflection point of the chiral condensate as a function of temperature.
One can also define the transition region
by the temperature range 
where 
$M$ varies between $0.4$ and $0.6$. This range depends on 
the magnetic field and gives rise to a band in the $B$--$T$ plane.
This is shown as the dark band in Fig.~\ref{physpoint}.
In the same manner, we define the crossover transition for the
deconfinement transition by the the temperature range where 
$0.4<\Phi<0.6$. This is shown as the light band in Fig.~\ref{physpoint}.
Moreover, as a guide to the eye, we also show the lines where
$M/M_{0}={1\over2}$ and $\Phi={1\over2}$, respectively.
For comparison, we also show the pseudocritical temperature for
the chiral transition in the NJL model (dashed line).
The curves indicate that the two transitions coincide
for magnetic fields up to $|qB|\approx 5M_{\pi}^2$ and after that they split.
The deconfinement transition is always taking place first.
We notice that $T_c$ for the chiral transition first decreases as a 
function of $|qB|$ until $|qB|\approx3M_{\pi}^2$ and then it starts to
increase again.
At the physical point where there is only a 
cross-over, the width of the transition is much bigger than this decrease in 
$T_{c}$. The difference between $T_c(|qB|=0)$ and $T_c(|qB|=3M_{\pi}^2)$
is only a few MeV which is comparable to the drop found by 
Bali et al.~\cite{budaleik,gunnar} for $N_c=3$.
However, note that these authors find that $T_c$ is decreasing as a function
in the entire region they investigated, namely magnetic fields up to
$|qB|\approx$ 1 (GeV)$^2$,
and that the drop in $T_c$ is approximately 20 MeV in this range.
The non-monotomic behavior of $T_c$ as a function of $|qB|$ in the PNJL
model is absent in the NJL model and we attribute this to the coupling to
the Polyakov loop. Another possible explanation for the non-monotonic behaviour 
is that it could be very well an artifact of our mean field approximation.
Finally, we note that the band of the chiral transition
is much narrower than that of the deconfinement transition 
(approximately $T=10$ MeV versus approximately $T=30$ MeV) and is so
significantly faster.
In Fig.~\ref{s1}, we show the normalized thermodynamic potential 
$\Omega/F_{\pi}^4$ as a function of $M$ and $\Phi$ 
in the chiral limit with $|qB|=20M_{\pi}^2$
for different temperatures. From left to right $T=100$ MeV,
$T=200$ MeV, $T=219$ MeV, and $T=250$ MeV.
For $T=100$ MeV, we are in the confined and the chirally
broken phase, while for $T=200$ MeV, the value of $\Phi$
is slightly above ${1\over2}$ ($\Phi={1\over2}$ for $T=197$ MeV).
The minimum is still for nonzero $M$ and so chiral symmetry is broken.
For $T=219$ MeV, $M/M_0$ is still nonozero ($T_c=220$ MeV)
and so we are just below the chiral transition.

\newpage

\section{Concluding remarks}
In the present paper, we have considered two-color two-flavor QCD
in a constant magnetic background using the PNJL model.
In the chiral limit, the chiral transition is always second order with
mean-field critical exponents. The order of the transition is in agreement with
universality arguments. 
Our results for the chiral transition as a function of $|qB|$
shows some interesting behavior:
the critical temperature
first decreases and then it increases.
This behavior is more pronounced at the physical point
(cf. Figs.~\ref{tctrans} and~\ref{physpoint}). 
In this case the transition temperature decreases 
for values of $|qB|$
up to approximately 3$M_{\pi}^2$
and then it increases.
We therefore have inverse magnetic catalysis in this range of $|qB|$
and results from the coupling to the Polyakov loop since
the transition temperature is increasing for all values of $|qB|$
in the NJL model. Even though we attribute the coupling to the Polyakov loop 
as the mechanism responsible for the inverse catalysis in the range mentioned, 
there is the possibility that this effect is actually an artifact of the 
mean-field approximation we use in our calculations and it would not survive if
we used more advanced techniques. For example it is known that critical points 
in the $\mu-T$ plane in some low-energy effective theories display a critical 
endpoint in mean-field calculations. If one uses functional renormalization 
techniques, they disappear~\cite{strod} . Thus, it is worthwhile pursuing this 
behaviour using more sophisticated methods before we conclude more firmly about 
the nature of this effect. The lattice simulations of Ref.~\cite{peter}
seem to indicate a transition temperature which is increasing
with the strength of the magnetic field.
However, one must be cautious since they did not take the 
continuum limit and used $N_f=4$ with identical electric charges.
A careful study using 1+1 flavors, $N_c=2$, and taking the continuum
limit is necessary to compare with the results in the present paper.

We next turn to the case $N_c=3$.
The inverse catalysis for temperature around $T_c$
seen in~\cite{budaleik,gunnar,falk,gunnar2}
hinges on taking continuum limit and using physical
quark masses. For larger unphysical quark masses, 
the system shows catalysis at finite temperatures.
Given the lattice results for $N_c=3$ it is clear that
all the model calculations to date fail at temperatures $T$
around $T_c$ and in particular does not incorporate
that magnetic catalysis is turned into inverse magnetic catalysis
around the transition. This is independent of whether it is a
mean-field calculation or one goes beyond using e.g. functional renormalization
group techniques~\cite{skokov1,anders}.
In the recent papers~\cite{falk,gunnar2}, the authors provide
a plausible explanation for the discrepancy between the
model calculations and the lattice simulations.
The chiral condensate can be written as
\begin{widetext}
\bqa
\langle\bar{\psi}\psi\rangle
&=&{1\over{\cal Z}(B)}\int d{\cal U}
e^{-S_g}\det(D\!\!\!\!/(B)+m){\rm Tr}
(D\!\!\!\!/(B)+m)^{-1}\;,
\label{det}
\eqa
where the partition function is
\bqa
{\cal Z}(B)&=&\int d{\cal U}
e^{-S_g}\det(D\!\!\!\!/(B)+m)\;,
\eqa
and $S_g$ is the pure-glue action. Thus there are two contributions to the
chiral condensate, namely the operator itself 
(coined valence contribution)
and the change of typical gauge configurations sampled, coming
from the determinant in Eq.~(\ref{det})
(coined sea contribution). At least for
small magnetic fields one can disentangle these contributions by 
defining
\bqa
\langle\bar{\psi}\psi\rangle^{\rm val}
&=&{1\over{\cal Z}(0)}\int d{\cal U}
e^{-S_g}\det(D\!\!\!\!/(0)+m){\rm Tr}
(D\!\!\!\!/(B)+m)^{-1}\;,
\\
\langle\bar{\psi}\psi\rangle^{\rm sea}
&=&{1\over{\cal Z}(B)}\int d{\cal U}
e^{-S_g}\det(D\!\!\!\!/(B)+m){\rm Tr}
(D\!\!\!\!/(0)+m)^{-1}\;.
\eqa
\end{widetext}
At zero temperature, both contributions are positive leading
to magnetic catalysis. At temperatures around the transitions temperature,
the valence condensate is still positive while the sea condensate
is negative. Hence there is a competition between the two leading to a
net inverse catalysis. The sea contribution can be viewed as a back reaction
of the fermions on the gauge fields and this effect is not present in the 
model calculations as there are no dynamical gauge fields.
If such a back reaction can be mimicked or 
incorporated in the model calculations,
one may be able to obtain agreement with the lattice simulations.

It is also interesting to note that the magnetic field hardly affects
the critical temperature for deconfinement. This is in line
with the observation of~\cite{mizher}.
Morever, our results
seem to indicate that the two transitions coincide at the physical
point up to fairly large values of the magnetic field.
$|qB|\approx 5M_{\pi}^2$.

Finally, we would like to comment on the role of quantum fluctuations
and related renormalization issues.
In a one-loop calculation of the effective potential it is possible
to separate the vacuum contributions from the
thermal contributions. In some case, it therefore makes sense to
investigate the role of the vacuum fluctuations. For example, in the
QM, it is customary to treat the 
bosons at tree level and the fermions at the one-loop level.
In this case it was shown in Ref.~\cite{skokov} for $B=0$
that the order
of the phase transition depends whether the zero-temperature
fluctuations are included or not; if they are, the chiral transition
is second order and if they are not, it is first order. 
The same effect of the vacuum fluctuations were found in the entire
$\mu_B$--$T$ plane 
in strong magnetic fields in~\cite{rashid}.
In contrast to the QM model with quarks, 
this question does not make sense in the (P)NJL model.
The reason is that chiral symmetry breaking 
in the (P)NJL
is always a loop effect in contrast to the QM model where it is
built into the tree-level potential. 
In a similar manner, Ref.~\cite{boomsma} finds that a crossover
transition (for $N_c=3$) at the physical point remains a crossover
at finite magnetic field $B$ in an NJL model calculation.
This is in contrast to Ref.~\cite{mizher} 
where it is found that strong magnetic fields turn the crossover
into a first-order transition.
In their work, the authors use the QM model and renormalize by
subtracting
the fermionic vacuum fluctuations at $B=0$. 

\section*{Acknowledgments}
J. O. A. would like to thank Tomas Brauner for useful discussions.


\begin{thebibliography}{99}

\bibitem{book}
Lect. Notes Phys. "Strongly interacting matter in magnetic fields" (Springer), 
edited by D. Kharzeev, K. Landsteiner, A. Schmitt, and H.-U. Yee.


\bibitem{duncan}R. C. Duncan and C. Thompson, Astrophys. J. 
{\bf 392} L9 (1992).

\bibitem{mag1}V. Skokov, A. Y. Illarionov, and V. Toneev, 
Int. J. Mod.Phys. A {\bf 24} 5025, (2009). 
 
\bibitem {mag2}A. Bzdak and V. Skokov, 
Phys.Lett. B {\bf 710}, 171 (2012).



\bibitem{warringa}D. E. Kharzeev, L. D. McLerran, and 
H. J. Warringa, Nucl. Phys. A {\bf 803}, 227 (2008).





\bibitem{vaksa}T. Vachaspati, Phys. Lett. B {\bf 265}, 258 (1991).
\bibitem{olesen}
K. Enqvist and P. Olesen, Phys. Lett. B {\bf 319}, 178 (1993).

\bibitem{laine}M. Laine
K. Kajantie, M. Laine, J. Peisa, K. Rummukainen, 
M. E. Shaposhnikov,
Nucl. Phys. B {\bf 544}, 357 (1999).


\bibitem{spanish}
A. De Simone, G. Nardini, M. Quiros, and A. Riotto,
JCAP 1110, 030 (2011).


\bibitem{klevansky}	
S. P. Klevansky and R. H. Lemmer, Phys. Rev. D {\bf 39}, 3478, (1989).

\bibitem{klimenko}
K.G. Klimenko, Z. Phys. C {\bf 54}, 323 (1992).

\bibitem{gusynin1}
V. P. Gusynin, V. A. Miransky, and I. A. Shovkovy,
Phys. Rev. Lett. {\bf 73}, 3499 (1994).

\bibitem{gusynin2}
V. P. Gusynin, V.A. Miransky, and I. A. Shovkovy, Nucl.
Phys. B {\bf 462}, 249 (1996).

\bibitem{ebert22}	
D. Ebert, K. G. Klimenko, M. A. Vdovichenko, and A. S. Vshivtsev,
Phys. Rev. D {\bf 61} 025005 (1999).


\bibitem{shushp}
I. Shushpanov and A. V. Smilga, Phys. Lett. B {\bf 402} 351, (1997).


\bibitem{werbos}
T. D. Cohen, D. A. McGady, and E. S. Werbos, Phys. Rev. C {\bf 76}, 
055201 (2007).

\bibitem{janm}K. Fukushima and J. M. Pawlowski,
Phys. Rev. D {\bf 86}, 076013 (2012). 


\bibitem{hidaka} 
K. Fukushima and Y. Hidaka, 
Phys. Rev. Lett., {\bf 110}, 031601 (2013).

\bibitem{quench}
V. V. Braguta, P. V. Buividovich, T. Kalaydzhyan, S. V. Kuznetsov, 
and M. I. Polikarpov, PoS LATTICE2010, 190 (2010), 
Phys. Atom. Nucl. {\bf 75}, 488 (2012).






\bibitem{budaleik}
G. S. Bali, F. Bruckmann, G. Endr\H{o}di, Z. Fodor, S. D. Katz, S. Krieg, 
A. Schafer, and K. K. Szabo,
JHEP {\bf 1202}, 044 (2012).

\bibitem{gunnar} 
G. S. Bali, F. Bruckmann, G. Endr\H{o}di, Z. Fodor, S.D. Katz, and 
A. Schafer,
Phys. Rev. D {\bf 86}, 071502 (2012)





\bibitem{gatto1} R. Gatto and M. Ruggieri, Phys. Rev. D {\bf 82}, 054027
(2010).

\bibitem{duarte}
D. C. Duarte, R. L. S. Farias, and R. O. Ramos,
Phys. Rev. D {\bf 84}, 083525 (2011).
	
\bibitem{agassi}
N. O. Agasian, Phys. Lett. B {\bf 488}, 39 (2000).


\bibitem{jensoa}
J. O. Andersen, Phys. Rev. D {\bf 86}, 025020 (2012);
JHEP {\bf 1210}, 005 (2012).



\bibitem{fedo}
N. O. Agasian and S. M. Fedorov, Phys. Lett. B {\bf 663}, 445 (2008).

\bibitem{sid}
S. S. Avancini, D. P. Menezes, M. B. Pinto, and C. Providencia,
Phys.Rev. D {\bf 85} 091901 (2012).

\bibitem{pintotc}
G. N. Ferrari, A. F. Garcia, and M. B. Pinto, 
Phys. Rev. D {\bf 86}, 096005 (2012).



\bibitem{fuku}
K. Fukushima, M. Ruggieri, and R. Gatto, Phys. Rev. D
{\bf 81}, 114031 (2010).


\bibitem{mizher}
E. S. Fraga and A. J. Mizher, Phys. Rev. D. {\bf 78}, 025016 (2008).



\bibitem{skokov1}
V. Skokov,
Phys. Rev. D {\bf 85}, 03426 (2012).

\bibitem{anders}
J. O. Andersen and A. Tranberg, 
JHEP {\bf 1208},  002 (2012).


\bibitem{rashid}
J. O. Andersen and R. Khan, 
Phys. Rev. D {\bf 85}, 065026 (2012).

\bibitem{scoop}
M. Ferreira, P. Costa, D. P. Menezes, C. Providencia, N. Scoccola,
arXiv:1305.4751v1  [hep-ph].

\bibitem{gatto2}
R. Gatto and M. Ruggieri, Phys. Rev. D {\bf 83}, 034016 (2011).

	




\bibitem{bag}
E. S. Fraga and L. F. Palhares, Phys. Rev. D {\bf 86}, 016008 (2012).


\bibitem{mizher2}
A. J. Mizher, M. N. Chernodub and E. S. Fraga, Phys.
Rev. D {\bf 82}, 105016 (2010).


\bibitem{largen}
E. S. Fraga, J. Noronha, and L. F. Palhares, 
Phys. Rev. D {\bf 87}, 114014 (2013).








\bibitem{sanf}
M. D'Elia, S. Mukherjee, and F. Sanfilippo,
Phys. Rev. D  {\bf 82}, 051501 R (2010).



\bibitem{negro}
M. D'Elia and F. Negro, 
Phys. Rev. D {\bf 83}, 114028 (2011).





\bibitem{inverse1}
T. Inagaki, D. Kimura, and T. Murata, 
Prog. Theor. Phys. {\bf 111}, 371 (2004).

\bibitem{inverse2}
F. Preis, Anton Rebhan, and Andreas Schmitt.
JHEP {\bf 1103}, 033 (2011).

\bibitem{gunnar2}
G. S. Bali, F. Bruckmann, G. Endrodi, F. Gruber, and A. Schaefer
JHEP {\bf 1304} 130 (2013).

\bibitem{falk}
F. Bruckmann, G. Endrodi, and T. G. Kovacs,
JHEP {\bf 1304}, 112 (2013).




\bibitem{kond}
L. A. Kondratyuk and M. M. Gianinni, 
Phys. Lett. B {\bf 269} 139 (1991).






\bibitem{kogut}
J. B. Kogut, M. A. Stephanov, and D. Toublan,
Phys. Lett. B {\bf 464}, 183 (1999).

\bibitem{kim}	
K. Splittorff, D.T. Son, and M. A. Stephanov, 
Phys. Rev. D {\bf 64}, 016003 (2001).

\bibitem{rotta}C. Ratti and W. Weise, Phys. Rev. D {\bf 70}, 054013 (2004).

\bibitem{cea}
P. Cea, L. Cosmai, M. D'Elia, and A. Papa 
JHEP {\bf 0702}, 066 (2007.)

\bibitem{simon}
S. Hands, S. Kim, J-I. Skullerud, Phys. Rev. D {\bf 81} 091502 (2010).

\bibitem{tilo}
T. Kanazawa, Tilo Wettig, N. Yamamoto,
JHEP {\bf 0908}, 003 (2009).


\bibitem{tomas1}
T. Brauner, K. Fukushima, and Y. Hidaka 
Phys. Rev. D {\bf 80}, 074035  (2009); Erratum-ibid. D {\bf 81}, 119904
(2010).


	
\bibitem{jens}
J. O. Andersen and T. Brauner,
Phys. Rev. D {\bf 81}, 096004 (2010).


\bibitem{zhang}
T. Zhang, T. Brauner, A. Kurkela, A. Vuorinen,
JHEP {\bf 1202}, 139 (2012).


\bibitem{strod}
N. Strodthoff, B.-J. Schaefer, L. von Smekal,
Phys. Rev. D {\bf 85}, 074007 (2012).

\bibitem{kashiw}K. Kashiwa, T. Sasaki, and H. Kounu,
Phys. Rev. D {\bf 87} 016015 (2013).


\bibitem{wiese}
S. Imai, H. Toki, and W. Weise
e-Print: arXiv:1210.1307 [nucl-th].





\bibitem{poli1}
P. V. Buividovich, M. N. Chernodub, E. V. Luschevskaya, and
M. I. Polikarpov, Phys. Lett. B {\bf 682}, 484 (2010).

\bibitem{poli2}
P. V. Buividovich, M. N. Chernodub, E. V. Luschevskaya, and
M. I. Polikarpov, Nucl. Phys. B {\bf 826}, 313 (2010).







\bibitem{peter}
E.-M. Ilgenfritz, M. Kalinowski, M. M\"uller-Preussker, B. Petersson, and
A. Schreiber, Phys. Rev. D {\bf 85}, 114504 (2012).


\bibitem{svit}	
B. Svetitsky and L. G. Yaffe, Nucl. Phys. B {\bf 210}, 423 (1982).

\bibitem{fukupol}
K. Fukushima, Phys. Lett. B {\bf 591}, 277 (2004).
	


\bibitem{megias}
E. Megias, E. Ruiz Arriola, and L.L. Salcedo,
Phys. Rev. D {\bf 74}, (2006) 065005; ibid {\bf 74}, 114014 (2006). 


\bibitem{he}
G.-F. Sun, L. He, P. Zhuang,  Phys. Rev. D {\bf 75}, 096004 (2007).



\bibitem{pure}C. Sasaki, B. Friman, and K. Redlich,
Phys. Rev. D {\bf 75}, 074013 (2007).

\bibitem{lucini}B. Lucini, A. Rago, and E. Rinaldi, Phys. Lett. B {\bf 712} 
279 (2012).

\bibitem{latfuk}K. Fukushima, Phys. Rev. D {\bf 77}, 114028 (2008).

\bibitem{frasca}
M. Frasca and  M. Ruggieri, Phys. Rev. D {\bf 83}, 094024 (2011).

\bibitem{calle}
N. Callebaut, D. Dudal, and H. Verschelde, PoS FACESQCD 046 (2010).

\bibitem{skokov}
V. Skokov, B. Friman, E. Nakano, K. Redlich, and 
B.-J. Schaefer, Phys.Rev. D {\bf 82}, 034029 (2010).

\bibitem{boomsma}
J. K. Boomsma and D. Boer, 
Phys. Rev. D {\bf 81}, 074005 (2010).

\end{thebibliography}
\end{document}